\documentclass[11pt]{article}
\usepackage{ascmac,amsmath}
\usepackage{bm}
\usepackage{graphicx,epsfig}
\usepackage{geometry}
\makeatletter
\renewenvironment{thebibliography}[1]
{\section*{\refname\@mkboth{\refname}{\refname}}%
  \list{\@biblabel{\@arabic\c@enumiv}}%
       {\settowidth\labelwidth{\@biblabel{#1}}%
        \leftmargin\labelwidth
        \advance\leftmargin\labelsep
 \setlength\baselineskip{11pt}%
        \@openbib@code
        \usecounter{enumiv}%
        \let\p@enumiv\@empty
        \renewcommand\theenumiv{\@arabic\c@enumiv}}%
  \sloppy
  \clubpenalty4000
  \@clubpenalty\clubpenalty
  \widowpenalty4000%
  \sfcode`\.\@m}
 {\def\@noitemerr
 {\@latex@warning{Empty `thebibliography' environment}}%
\endlist}
\makeatother
\begin{document}
\centerline{{\sl Genshikaku Kenkyu Suppl.} No. 000 (2012)}
\begin{center} 
\vskip 2mm
{\Large\bf

Search of Deeply Bound Kaon States at B-factories
\hspace{-1mm}\footnote{Presented at the International Workshop on Strangeness 
Nuclear Physics (SNP12), August 27 - 29, 2012, \\
\hspace*{5mm} Neyagawa, Osaka, Japan.}
}\vspace{5mm}

{
Simonetta Marcello$^{a, b}$, Francesca De Mori$^{a, b}$ 
and Alessandra Filippi$^b$
}\bigskip

{\small
$^a$Department of Physics, Torino University, via P. Giuria 1, 
Torino 10125, Italy\\ 
$^b$INFN Sezione di Torino, via P. Giuria 1, Torino 10125, Italy\\ 
}
\end{center}
\vspace{3mm}

\noindent
{\small \textbf{Abstract}:\quad
The study of deeply  bound kaon states is usually carried out or proposed 
using $K^-$ or $\pi^+$ beams with fixed nuclear target or proton proton interactions
or, more recently, heavy ion collisions.
The extension of the measurements to different and new environments could allow
to clarify their possible existence.
The high luminosity of 10$^{36}$cm$^{-2}$s$^{-1}$, 
which will be achieved at future B-factories, both in Italy and in Japan, 
opens the possibility to search for deeply bound kaon states in the vacuum, 
in the glue-rich $\Upsilon(nS)$ decays.
Here the opportunity to perform a measurement of the simplest kaonic nuclear state,
the $K^-pp$ and other bound states, at B-factories is discussed.

}%


\section{Introduction}

Nowadays, the study of $\bar{K}$-nuclear bound states is of great interest in nuclear hadron physics.
After the prediction of these states by Akaishi and Yamazaki \cite{DBKS},
several theoretical calculations have been done and different experiments have been proposed. 
In their model Akaishi and Yamazaki used a phenomenological $\bar{K}-N$ potential, 
based on experimental data of free $\bar{K}-N$ scattering lengths
and of kaonic hydrogen X-rays, with the peculiar ansatz
that the $\Lambda(1405)$ resonance ($\Lambda^*$ in the following) 
is a $\bar{K}-N$ quasi-bound state embedded in a continuum of $\Sigma \pi$, 
instead of a simple baryon with a three-quark structure,
and it behaves like a seed which can cluster one or more nucleons, 
producing a $\bar{K}$ nuclear bound state.
In this model the strongly attractive $\bar{K}-N$ interaction, in the 
isospin $I = 0$ state, may lead to tremendously 
condensed systems with a cluster structure and a central nucleon density 
as high as 1.5 $\it nucleons$/fm$^3$. Therefore, in this scenario very narrow ($<$ 100 MeV) 
and deeply bound ($B \sim100$ MeV) kaon states (DBKS), 
such as $K^-pp$, $K^-pn$, $K^-ppp$, $K^-ppn$, $K^-pppn$ and $K^-ppnn$, 
are expected. 
Due to their large binding energy the main decay channel to $\Sigma \pi$ is energetically
forbidden.
The simplest nuclear DBKS, the $K^-pp$, is predicted to be a quasi stable state 
with a mass $M = 2322$ MeV/c$^2$, a binding energy $B = 48$ MeV and a partial decay
width $\Gamma$~=~61 MeV.

Several theoretical calculations \cite{Theory-DBKS}, based on different approaches,
have been developed and their results are quite in disagreement,
even if they do not deny the existence of these states.
The main difference concerns the potential, whether it is shallow or deep 
and, then, if the decay width is narrow enough, compared to the binding energy,
to allow their observation. 
One more problem comes from the Final State Interactions (FSI) which could fake a signal \cite{Magas}, 
so that positive experimental results could be attributed to such
an effect.

Since DBKS are very dense systems, made of a strange particle 
and a few nucleons, the study of this system 
is related to kaon condensation \cite{KCond} and 
is important to understand the features of 
astrophysical objects such as the neutron stars, where the nuclear density
could reach ten times the ordinary one and where strangeness degrees of freedom
could play a crucial role \cite{Weber}. This could lead to the formation 
of strange quark matter, which could be more stable than the ordinary 
nuclear matter. 
For this reason, even after some years, the study of DBKS keeps on being 
a hot topic in hadron physics.

The FINUDA experiment \cite{FINUDA} at the DA$\Phi$NE $e^+e^-$ 
collider in Frascati, claimed for the first time the observation of
the lightest nuclear DBKS, the $K^-pp$, 
using a low energy $K^-$ beam stopped in very thin nuclear targets. 
A peak ($\Gamma \sim 70$ MeV) was 
observed in the $\Lambda$p invariant mass spectrum 
at a mass $\sim$2.26 GeV/c$^2$ corresponding to a binding energy 
B $\sim 115$ MeV. 
Therefore, the $\Lambda$p pairs,
which exhibited a clean back-to-back topology,  
were interpreted as the decay products of such a state.

In recent analyses of old experiments, such as OBELIX \cite{Obelix} at CERN 
with $\bar{p}-^4He$ annihilations at rest
and DISTO \cite{DISTO} at SATURNE with the exclusive 
$pp \rightarrow p K^+ \Lambda$ production at $T_p$ = 2.85 GeV, 
hints of DBKS have been also found. 
Nevertheless, their existence has not been established yet 
and further measurements 
are planned in the near future at several facilities. Since all of them
use $K^-$ or $\pi^+$ beams interacting with nuclear targets or heavy ion 
collisions, where the formation of the nuclear kaonic state occurs 
in the nuclear medium, new experiments in different environments 
could help to find out signatures of the existence of such states. 
In particular, a search of production of DBKS in the vacuum, 
rather than in the medium, can be performed at B-factories,  
in the glue-rich decays of the lightest bottomonia.

\section{Production of baryons and deuteron 
in $\Upsilon(1S)$ and $\Upsilon(2S)$ decays}

\subsection{Baryon production}
In the bottomonium spectrum below the $B\bar{B}$ threshold 
there are three states, 
the $\Upsilon(1S)$, $\Upsilon(2S)$ and $\Upsilon(3S)$. 
The large mass of these resonances ($\sim 10$ GeV/c$^2$) is suitable to produce 
through their decay baryons and anti-baryons, and few-body bound systems, as well.
It should be noted that the branching ratio for the strong decays 
of $\Upsilon(1S)$ amounts to $\sim80\%$. 
These direct decays proceed via three gluons, hence the decay
products are the result of their hadronization.
These glue-rich decays might also produce exotic multiquark states. 
For the $\Upsilon(2S)$ and $\Upsilon(3S)$ resonances, direct radiative decays,
both e.m. and hadronic, compete with the gluon annihilation modes;
but these states can be also used to produce the $\Upsilon(1S)$.

It is well known, since many years, that baryon production from $\Upsilon$(1S) 
is enhanced with respect to the continuum hadronization \cite{Behrends}. 
In particular, the CLEO experiment \cite{BaryonCLEO} at CESR, 
studying the inclusive 
production of baryons/antibaryons in $e^+e^-$ collisions 
at $\sqrt{s} \sim 10$ GeV, 
finds out that the enhancements of per-event total particle yields 
for the $\Lambda$ hyperon, proton and antiproton are about a factor of two  
in the $\Upsilon (1S) \rightarrow ggg$ decays as compared to the 
non resonant nearby $q\bar{q}$ continuum ($e^+e^- \rightarrow q\bar{q}$).
The enhancement for $\Lambda$ production is the highest one ($\sim 2.7$). 
These results are not accounted for by the JETSET 7.3 fragmentation model. 
For the decays through gg$\gamma$ the observed enhancements 
are smaller (except for $\Lambda$'s, being in this case a factor of $\sim$ 2) 
and they are more in agreement with JETSET expectations.

\subsection{Deuteron production}

Most interesting, concerning the differences between
quark/gluon-fragmentation in the $\sqrt{s}$~$\sim 10$~GeV energy range, 
is the experimental evidence of bound state production, 
such as deuteron ($d$) and antideuteron ($\bar{d}$),
in gluonic decays of $\Upsilon$(1S) and $\Upsilon$(2S), 
that was first observed by ARGUS \cite{ARGUS} with very low statistics and, 
recently, by CLEO \cite{AntiD-CLEO}.
In particular, CLEO finds out that the branching ratio 
$\cal{B}$($\Upsilon (1S) \rightarrow \bar{d}X$)
is of the order of 3$\times$10$^{-5}$. 
Therefore, the results show
that $\bar{d}$ production is at least three times more likely in 
the hadronization of $\Upsilon (nS)$ (for n = 1, 2) through ggg and gg$\gamma$, 
compared to the hadronization of $q\bar{q}$ for the continuum production, 
where only an upper limit of 1$\times$10$^{-5}$ 
has been evaluated from the 
$e^+e^- \rightarrow \bar{d} X$ cross section. 
Besides, an upper limit of 1$\times$10$^{-5}$ has been 
measured for the production of $\bar{d}$ from $\Upsilon(4S)$.
 
CLEO also investigates how often the $\bar{d}$ baryon number 
is compensated by a $d$ or by a combination of two nucleons (n, p): 
the first case occurs 1\% of the times.  
Since the theoretical description of $d$ or $\bar{d}$ formation is based 
on the coalescence model \cite{Coalescence}, where an $\bar{n}$ 
and an $\bar{p}$ close to each other in phase 
space bind together, a precise determination of such a figure is important,
because double coalescence is unlikely and a different production 
mechanism, such as a primary (globally) production might be involved.

It can be also interesting to note that in heavy ion collisions the 
formation of bound systems, 
such as light nuclei/anti-nuclei or hypernuclei/anti-hypernuclei, 
occurs through coalescence processes. For instance, 
the STAR \cite{STAR-antihyp} 
experiment at RHIC has measured the ratios of the yields of 
anti-$^3_{\Lambda}H$ and $^3_{\Lambda}H$ and of anti-$^3He$ and $^3He$ 
in Au-Au collisions
at $\sqrt{s_{NN}}$ = 200 GeV, which amount to $\sim$ 0.5,
favouring the coalescence hypothesis. At STAR (anti)hypernuclei are produced
with similar yields of (anti)nuclei, different from what happens 
at lower energies.
Recently, the FOPI experiment \cite{FOPI} at GSI has started to search for DBKS 
in  Ni+Ni and Al+Al collisions at 1.93 AGeV, but their identification in the medium 
is however a challenge.

\section{Search of deeply bound kaon states at B-factories}

Since bound systems, such as antideuterons, have been 
found in the decays of $\Upsilon(nS)$, with n~=~1,~2, 
we have proposed \cite{marcello-FB20} to search for DBKS 
at the future SuperB factory \cite{SUPERB}, 
projected to be built in the Tor Vergata University Campus in Rome. 
Here, large numbers of heavy leptons and heavy quark mesons 
will be produced using an e$^+$e$^-$ asymmetric collider, operated at 
a c.m. energy corresponding to the rest mass of the bottomonium 
resonance $\Upsilon(4S)$ with a luminosity of 10$^{36}$cm$^{-2}$s$^{-1}$, 
looking mainly for new physics.  
A magnetic spectrometer with a large solid angle, designed to detect 
and fully reconstruct both charged and neutral particles with high efficiency 
and energy resolution, will be used.
Operating the machine at an energy below the $B\bar{B}$ 
threshold it will be possible to select the rest mass of the lightest bottomonium resonances
$\Upsilon$(nS) (with n = 1, 2, 3), in order to study the production of DBKS
in the$\Upsilon$ decays which proceed through the hadronization 
of three gluons. 
The same study can be performed with the Belle II detector \cite{BELLE-II}
at the future SuperKEKB facility in Japan, where a high 
luminosity comparable with the SuperB one will be available.
In fact, thanks to the nano-beam scheme, devised at LNF-INFN, an unprecedented 
integrated luminosity of 10~$ab^{-1}$/year will be achieved at both the machines,
giving the possibility to study very rare decays.

The lightest DBKS, the $K^-$pp, 
can be identified through its $\Lambda p$
decay mode, searching for an enhancement with $\Gamma<$ 100 MeV at a mass 
of about 2.3 GeV/c$^2$ 
in the $\Lambda p$ invariant mass spectrum and 
measuring the $\Lambda-p$ angular correlations, which can give imporant 
hints on the nature of the event. 
We can rely on the high performance of the apparata at B-factories 
for particle ID and full topological event reconstruction for an effective
identification of the DBKS.

Since there is no surrounding medium, FSI are negligible with respect to 
experiments using kaon beams on nuclear targets or heavy ion 
collisions. Moreover, we expect that the identification in the vacuum
will be easier and cleaner than in heavy ion collisions.

The first step to search for DBKS is to measure the production of $\Lambda^*$ 
in $\Upsilon(nS)$ decays, since it is the doorway particle 
through which it is possible to create these exotic states.
This resonance can be considered as a quasi-bound state of meson and baryon, dynamically
generated from the meson baryon interaction in coupled channels. 
It would be interesting to compare the production yields of this excited state 
in $\Upsilon(nS)$ decays with those in the $q\bar{q}$ continuum. 

In particular, the $\Lambda^*$ is decsribed by the superimposition 
of two $I = 0$ poles \cite{Lambda*}: 
one with a small width ($\Gamma \sim 30$ MeV) at the energy of $1420$ MeV, 
which couples mostly to $\bar{K}N$, and the other one with a larger width 
(120-250 MeV, depending on the model) at the energy of $1395$ MeV, 
which couples to $\Sigma \pi$. 
Since $\Lambda^*$ is under the $\bar{K}N$ threshold 
it decays about 100\% into the $\Sigma \pi$ channel. 
Unfortunately, it largely overlaps with the nearby $\Sigma^0(1385)$ excited state, therefore, 
in order to disantangle the $\Lambda^*$ from the $\Sigma^0(1385)$, 
the decay channel into $\Sigma^0 \pi^0$ has to be selected \cite{ANKE}, 
since isospin conservation forbids it for the decay of $\Sigma^0(1385)$. 
Moreover, the high luminosity available at the future B-factories may allow 
also the measurement of the $\Lambda^*$ radiative decays $\Lambda \gamma$ and $\Sigma^0 \gamma$, 
which have branching ratios three orders of magnitude 
smaller than the main $\Sigma \pi$ channel. 
Indeed, these decays could be an important test of the $\Lambda^*$ pole structure 
and could shed light on its real nature. 

\section{Search of more exotic bound systems in the $\Upsilon(nS)$ decays}

The observation of antideuterons in the $\Upsilon(1S)$ and $\Upsilon(2S)$ 
decays opens also the possibility 
to search for other nuclear bound states. For instance, light hypernuclei and 
anti-hypernuclei, with a $\Lambda$ or $\bar{\Lambda}$ hyperon (Y) 
could be produced in the glue-rich decays of $\Upsilon(nS)$,
as suggested by Roberto Mussa \cite{Mussa}. In particular, 
the large mass of the lightest $\Upsilon(nS)$, $m \sim 10$ GeV/c$^2$, is suitable to produce 
$\Lambda$-hypernuclei, such as $^3_{\Lambda}H$, $^4_{\Lambda}H$, $^4_{\Lambda}He$, or 
the respective anti-hypernuclei. 
The purpose of this study is not hypernuclear spectroscopy, which can be studied 
much better with high statistics and high energy resolution by the coming 
experiments at J-PARC, but the measurement of the production yields, which could give 
important clues on the formation process of these bound states, allowing to determine 
whether the coalescence or a different mechanism is involved.
Of course, the higher is the numbers of baryons forming the bound state,
the lower is the expected production yield. But the measurement is affordable
with the luminosity which will be available at the future B-factories.

The formation of light charmed hypernuclei, such as the charmed deuteron,
$\Lambda_c N$, or the respective anti-$\Lambda_c N$, 
can be studied in the $\Upsilon(nS)$ decays as well. 
The possible existence of $\Lambda_c$- and $\Sigma_c$-hypernuclei was predicted 
more than 30 years ago \cite{cHyp-Dover}, using the SU(4) symmetry 
to extend the one boson exchange (OBE) model to the $NN$ and $YN$ interactions. 
Recently, new theoretical calculations \cite{cHyp-Oka} using the OBE model approach,
but based on heavy quark spin symmetry and chiral symmetry for light quarks, 
have been done. 
It has been found that molecular bound states of $\Lambda_c N$ could exist
with both $J^P = 0^+$ and $J^P = 1^+$, with binding energies of at most tens of MeV.
In this model the formation of $\Lambda_c N$ bound states occurs after 
the production of $\Lambda_c$ and through the coalescence mechanism, 
as it is in the molecular picture of the formation of loosely bound
hadronic molecules, such as the X(3872) \cite{Braaten}.
Among these charmed hypernuclei, if they exist, the anti-$\Lambda_c N$
are the easiest to be identified, due to the presence of an anti-Nucleon in the final 
decay products, which has a particular signature when it annihilates in the detector.

The formation of other exotic deeply bound states in the vacuum can be also investigated.
For instance, production of double-$\bar{K}$ clusters, $K^- K^- NN$, 
can be considered.  
The $K^-K^-pp$ system has been predicted \cite{KKpp-Yamazaki} to be deeply bound, 
as an extension of the $K^-pp$ system. This system should be very compact, 
more than $K^-pp$.
The formation mechanism could proceed through the creation of two $\Lambda^*$ 
(as doorway particles),
which stick together forming the bound system. 
Its binding energy has been calculated to be 
$B = 117$ MeV, larger than for $K^-pp$, and its width to be very narrow, 
$\Gamma = 35$ MeV. This deeply bound state is expected to decay according to  
$(K^-K^-pp) \rightarrow \Lambda \Lambda$,
therefore, it should be easily identified.
A different theoretical calculation \cite{KKpp-Barnea} for this system, 
based on chiral interaction models, 
provides a lower binding energy close to 30 MeV and a larger width of about 80 MeV,
making this state difficult to be experimentally observed.

Another interesting deeply quasi-bound state, similar to the $\bar{K}NN$ one, 
but with a charm quark instead of a strange one, is the D-meson nuclear cluster 
$DNN$, with isospin $I = 1/2$.
Theoretical calculations \cite{DNN}, based on variational methods and 
Faddeev-FCA (Fixed Centre Approximation) equations,
predict this system as a narrow quasi-bound state with 
a width $\Gamma < 40$ MeV (including both the mesonic decay and the D absorption) 
and a large binding energy, $B \sim 250$ MeV. 
The formation mechanism is based on the creation of a $\Lambda_c(2595)$ 
($\Lambda_c^*$ in the following), 
which plays the same role of the $\Lambda^*$ in the formation of DBKS.
The strong attractive interaction in $I = 0$ dynamically generates 
this $J^P = 1/2^-$ $\Lambda_c^*$ excited state, with the advantage to have 
a rather narrow width, $\Gamma < 1.9$ MeV (compared to the $\Lambda^*$),
which does not affect the width of the following $DNN$ quasi-bound state. 
A smaller binding, but still large enough, can be expected when this system
is produced via the coalescence  of $\Lambda_c^*N$. 
Such a $DNN$ quasi-bound state may be seen through the decay channels
$\Lambda_c N$ or $\Lambda_c \pi N$, and especially   
$\Lambda_c p$ or $\Lambda_c \pi^- p$ channels can be easily identified in the analysis 
of the respective invariant mass spectra.

An investigation of the production of $\Lambda_c^*$ excited state 
in $\Upsilon(nS)$ decays can be pursued
in an analogous way of the $\Lambda^*$.

\section{Conclusions}
\label{sec:3}
In the last decade the study of deeply bound kaon states has been, and still remain, 
a hot topic in hadron physics,
because their existence is related to kaon condensation and to the physics of 
the core of neutron stars.
Theoretical models are not in agreement about their existence 
and only scarce experimental measurements in the nuclear medium are available.
New experiments in different environments would be useful 
to compare the results.  
Search of the light $K^-pp$ state can be extended from the nuclear medium
to the vacuum, looking for its production in the glue-rich decays 
of the lightest $\Upsilon(nS)$ bottomonia at the future B-factories, 
taking advantage of their high luminosity.

Measurements of the production yields of this state from the 
lightest bottomonium should give important clues, 
not only about the nature of these states, but also 
about the hadronization processes of quarks and gluons. 

Moreover, a wider investigation can be attempted extending this study to the possible 
formation of other bound states, such as hypernuclei and charmed deuteron
and deeply bound states with strangeness $S = -2$ or charm $C = +1$.

\end{document}